\documentclass[11pt,twoside]{./atmp}

\usepackage{amsmath,amssymb}
\usepackage{cite}
\usepackage[all]{xy}
\usepackage{color}
\theoremstyle{plain}

\newtheorem{exmp}{Example}
\theoremstyle{definition}

\theoremstyle{remark}

\numberwithin{equation}{section}

\begin{document}

\title[DJM  for Laplace equation]
{The DJ method for exact solutions of Laplace equation}


\author[M. Yaseen et.al]{M. Yaseen, M. Samraiz, S. Naheed}

\address{Department of Mathematics, University of Sargodha, Sargodha, Pakistan.}  
\begin{abstract}
In this paper, the iterative method developed by Daftardar-Gejji and Jafari (DJ method) is employed for analytic treatment of Laplace equation with Dirichlet and Neumann boundary conditions. The method is demonstrated by several physical models of Laplace equation. The obtained results show that the present approach is highly accurate  and requires reduced amount of calculations compared with the existing iterative methods.
\end{abstract}
\maketitle
\section{Introduction}
The Daftardar-Jafari method (DJM) developed in 2006 has been extensively used by many researchers for the treatment of linear and nonlinear ordinary and partial differential equations of integer and fractional order \cite{varsha2,varsha3,DJ4,DJ5,DJ6,DJ7,DJ8}. An excellent modification of DJM can be found in \cite{yaseen9} for solving different models of linear and nonlinaer Klein-Gordon equations. The method converges to the exact solution if it exists through successive approximations. For concrete problems, a few number of approximations can be used for numerical purposes with high degree of accuracy. The DJM does not require any restrictive assumptions for nonlinear terms as required by some existing techniques. The aim of this work is to effectively employ DJM to obtain exact solutions for different models of  Laplace
equation. While the variational iteration method \cite{wazwaz1,He10,A.M11} requires the determination of Lagrange multiplier in its computational algorithm, DJM is independent of any such requirements. Moreover, unlike the Adomian decomposition method \cite{adomian12,kaya13,basak14,wazwaz15,wazwaz16,mkhan17,yusufogla18, Abdou19,zaid20}, where the calculation of the tedious Adomian polynomials is needed to deal with nonlienar terms, DJM handles linear and nonlinear terms in a simple and straightforward manner without any additional requirements.

As mentioned before, we aim to obtain exact solutions of four models of Laplace equation, two with Dirichlet boundary conditions and
two with Neumann boundary conditions. We also aim to establish that DJM is powerful, promising and efficient in handling nonlinear engineering problems. In what follows, we give a brief review of DJM.



\noindent

\section{The DJ Method}
Consider the following general functional equation
\begin{equation}  \label{2.1}
u(\overline{x})=f(\overline{x})+ N(u(\overline{x}))
\end{equation}
where $N$ is a nonlinear operator from a Banach space $B \rightarrow B$ and $%
f$ is a known function. $\overline{x}=(x_1,x_2,\cdots,x_n)$. We are looking
for a solution $u$ of (\ref{2.1})having the series form
\begin{equation}  \label{2.2}
u(\overline{x})=\sum\limits_{n=0}^\infty u_i(\overline{x})
\end{equation}
The nonlinear operator $N$ can be decomposed as
\begin{equation}  \label{2.3}
N(\sum\limits_{i=0}^\infty u_i)=N(u_0)+\sum\limits_{i=1}^\infty
\{N(\sum\limits_{j=0}^i)u_j-N(\sum\limits_{j=0}^{i-1} u_j)\}.
\end{equation}
From equations (\ref{2.2})and (\ref{2.3}), equation (\ref{2.1}) is
equivalent to
\begin{equation}  \label{2.4}
\sum\limits_{i=0}^\infty u_i=f+N(u_0)+\sum\limits_{i=1}^\infty
\{N(\sum\limits_{j=0}^i)u_j-N(\sum\limits_{j=0}^{i-1} u_j)\}
\end{equation}
We define the recurrence relation
\begin{eqnarray}
\left\{%
\begin{array}{l}
\label{1} u_0=f \\
u_1=N(u_0) \\
u_{m+1}=N(u_0+\cdots+u_m)-N(u_0+\cdots+u_{m-1})%
\end{array}%
\right.
\end{eqnarray}
Then
\begin{equation}  \label{2.6}
(u_1+\cdots+u_m)=N(u_0+\cdots+u_m)
\end{equation}
and
\begin{equation}  \label{2.7}
\sum\limits_{i=0}^\infty u_i=f+N(\sum\limits_{i=0}^\infty u_i)
\end{equation}
The $k-$term approximate solution of (\ref{2.1}) is given by $%
u=u_0+u_1+\cdots+u_{k-1}$
In what follows, we apply DJM to four physical models to demonstrate the strength of the method and to establish exact solutions of these models.
\begin{exmp}
Consider the second order laplace equation,
\begin{equation}
\ \ u_{xx}-u_{yy}=0,~~~~~~0<x,y<\pi  \label{3.1}
\end{equation}
with boundary conditions
\begin{eqnarray}\label{3.2}
\begin{array}{l}
u(x,0)=\sinh(x),~u(x,\pi)=-\sinh(x)\\
u(0,y)=0,~~~u(\pi,y)=\sinh(\pi)\cos(y)
\end{array}
\end{eqnarray}
\end{exmp}
The equation (\ref{3.1}) is equivalent to the following integral equation
\begin{equation}\nonumber
u=\sinh(x)+yg(x)-\int\limits_{0}^{y}\int\limits_{0}^{y}(u_{xx}(x,y))dydy
\end{equation}
where $ g(x)=u_{y}(x,0)$.\\
Set $u_{0}=\sinh(x)+yg(x)$ and $N(u)=-\int\limits_{0}^{y}\int\limits_{0}^{y}(u_{xx}(x,y))dydy.$\\
Following the algorithm (\ref{1}), the successive approximations are
\begin{eqnarray}
\begin{array}{l}\nonumber
u_{1}=N(u_{0})=-\frac{1}{2!}y^2\sinh(x)-\frac{1}{3!}y^3 g''(x)\\
u_{2}=N(u_{0}+u_{1})-N(u_{0})=\frac{1}{4!}y^4\sinh(x)+\frac{1}{5!}y^5g^{iv}(x)\\
u_{3}=N(u_{0}+u_{1}+u_{2})-N(u_{0}+u_{1})=-\frac{1}{6!}y^6\sinh(x)-\frac{1}{7!}y^7g^{vi}(x)\\
\vdots
\end{array}
\end{eqnarray}
Combining the above results obtained for the components yields
\begin{equation}\nonumber
u(x,y)=\sinh(x)+yg(x)-\frac{1}{2!}y^2\sinh(x)-\frac{1}{3!}y^3g''(x)+\frac{1}{4!}y^4\sinh(x)+\frac{1}{5!}y^5g^{iv}(x)+\cdots
\end{equation}
Using the inhomogeneous boundary condition, $u(\pi,y)=\sinh(\pi)\cos(y)$ and the Taylor expansion of $\cos(y)$, we obtain
$$\sinh(\pi)+yg(\pi)-\frac{1}{2}y^2\sinh(\pi)-\frac{1}{6}y^3g''(\pi)+\frac{1}{24}y^4\sinh(\pi)+\cdots=
\sinh(\pi)(1-\frac{y^2}{2!}+\frac{y^4}{4!}-\cdots)$$
Equating the coefficients of like terms on both sides gives
$$g(\pi)=g''(\pi)=g^{iv}(\pi)=\cdots=0$$
This means that $g(x)=0$. Consequently, the solution in series form is given by
\begin{equation}\nonumber
u(x,y)=\sinh(x)(1-\frac{y^2}{2!}+\frac{y^4}{4!}-\cdots)
\end{equation}
and in closed form $$u(x,y)=\sinh(x)\cos(y)$$ which is also the exact solution.
\begin{exmp}
Consider the second order laplace equation,
\begin{equation}
\ \ u_{xx}-u_{yy}=0,~~~~~~0<x,~~~y<\pi  \label{4.1}
\end{equation}%
with boundary conditions
\begin{eqnarray}\label{4.2}
\begin{array}{l}
u(x,0)=0,~~~u(x,\pi)=0\\
u(0,y)=\sin(y),u(\pi,y)=\cosh(\pi)\sin(y)
\end{array}
\end{eqnarray}
\end{exmp}
The equation (\ref{4.1}) is equivalent to the following integral equation
\begin{equation}\nonumber
u(x,y)=yg(x)-\int\limits_{0}^{y}\int\limits_{0}^{y}(u_{xx}(x,y))dydy
\end{equation}
where $g(x)=u_{y}(x,0)$.
Set $u_{0}=yg(x)$ and $N(u)=-\int\limits_{0}^{y}\int\limits_{0}^{y}(u_{xx}(x,y))dydy.$\\
Following the algorithm (\ref{1}), the successive approximations are
\begin{eqnarray}
\begin{array}{l}\nonumber
u_{1}=N(u_{0})=-\frac{1}{3!}y^3g''(x)\\
u_{2}=N(u_{0}+u_{1})-N(u_{0})=\frac{1}{5!}y^5g^{iv}(x)\\
u_{3}=N(u_{0}+u_{1}+u_{2})-N(u_{0}+u_{1})=-\frac{1}{7!}y^7g^{vi}(x)\\
\vdots
\end{array}
\end{eqnarray}
Combining the above results, we obtain
\begin{equation}\nonumber
u(x,y)=y g(x)-\frac{1}{3!}y^3g''(x)+\frac{1}{5!}y^5g^{iv}(x)-\frac{1}{7!}y^7g^{vi}(x)\cdots
\end{equation}
Using the inhomogeneous boundary condition, $u(\pi,y)=\cosh(\pi)\sin(y)$ and the Taylor expansion of $\sin(y)$, we obtain
$$y g(\pi)-\frac{1}{3!}y^3g''(\pi)+\frac{1}{5!}y^5g^{iv}(\pi)-\frac{1}{7!}y^7g^{vi}(\pi)+\cdots=
\cosh(\pi)(y-\frac{y^3}{3!}+\frac{y^5}{5!}-\cdots).$$
Equating the coefficients of like terms on both sides gives
$$g(\pi)=g''(\pi)=g^{iv}(\pi)=\cdots=\cosh(\pi)$$
This means that
$$g(x)=\cosh(x)$$
Consequently, the solution in series form is given by
\begin{equation}\nonumber
u(x,y)=\cosh(x)(y-\frac{y^3}{3!}+\frac{y^5}{5!}-\cdots)
\end{equation}
and in closed form $$u(x,y)=\cosh(x)\sin(y)$$ which is also the exact solution.
\begin{exmp}
Consider the second order laplace equation,
\begin{equation}
\ \ u_{xx}-u_{yy}=0,~~~~0<x,y<\pi  \label{5.1}
\end{equation}%
with boundary conditions
\begin{eqnarray}\label{5.2}
\begin{array}{l}
u_y(x,0)=0,~u_y(x,\pi)=2\cos(2x)\sinh(2\pi)\\
u_x(0,y)=0,~u_x(\pi,y)=0
\end{array}
\end{eqnarray}
\end{exmp}
The equation (\ref{5.1}) is equivalent to the following integral equation
\begin{equation}\nonumber
u=g(x)-\int\limits_{0}^{y}\int\limits_{0}^{y}(u_{xx}(x,y))dydy
\end{equation}
where $ g(x)=u(x,0)$.
Set $u_{0}=g(x)$ and $N(u)=-\int\limits_{0}^{y}\int\limits_{0}^{y}(u_{xx}(x,y))dydy.$\\
Following the algorithm (\ref{1}), the successive approximations are
\begin{eqnarray}
\begin{array}{l}\nonumber
u_{1}=N(u_{0})=-\frac{1}{2!}y^2g''(x)\\
u_{2}=N(u_{0}+u_{1})-N(u_{0})=\frac{1}{4!}y^4g^{iv}(x)\\
u_{3}=N(u_{0}+u_{1}+u_{2})-N(u_{0}+u_{1})=-\frac{1}{6!}y^6g^{vi}(x)\\
\vdots
\end{array}
\end{eqnarray}
Combining the above results, we obtain
\begin{equation}\label{5.3}
u(x,y)=g(x)-\frac{1}{2!}y^2g''(x)+\frac{1}{4!}y^4g^{iv}(x)-\frac{1}{6!}y^6g^{vi}(x)+\cdots
\end{equation}
Using the inhomogeneous boundary condition, $u_y(x,\pi)=2\cos(2x)\sinh(2\pi)$ and the Taylor expansion of $\sinh(2\pi)$, we obtain
$$-\pi g''(x)+\frac{1}{3!}\pi^3g^{iv}(x)-\frac{1}{5!}\pi^5g^{vi}(x)+\cdots=
2\cos(2x)(2\pi+\frac{8\pi^3}{3!}+\frac{32\pi^5}{5!}+\cdots)$$
Equating the coefficients of like terms on both sides gives
$$g''(x)=-4\cos(2x)$$
which gives
$$g(x)=\cos(2x)+C_0$$
Consequently, the solution in series form is given by
\begin{equation}\nonumber
u(x,y)=\cos(2x)(1+\frac{4y^2}{2!}+\frac{16y^4}{4!}+\cdots)+C_0
\end{equation}
and in closed form $$u(x,y)=\cos(2x)\cosh(2y)+C_0$$ which is also the exact solution.
\begin{exmp}
Consider the second order laplace equation,
\begin{equation}
\ \ u_{xx}-u_{yy}=0,~~~~~0<x,y<\pi  \label{6.1}
\end{equation}%
with boundary conditions
\begin{eqnarray}\label{6.2}
\begin{array}{l}
u_y(x,0)=\cos(x),~u_y(x,\pi)=\cosh(\pi)\cos(x)\\
u_x(0,y)=0,~~~~u_x(\pi,y)=0
\end{array}
\end{eqnarray}
\end{exmp}
The equation (\ref{6.1}) is equivalent to the following integral equation
\begin{equation}\nonumber
u=g(x)+y\cos(x)-\int\limits_{0}^{y}\int\limits_{0}^{y}(u_{xx}(x,y))dydy
\end{equation}
where $ g(x)=u(x,0)$.
Set $u_{0}=g(x)+y\cos(x)$ and $N(u)=-\int\limits_{0}^{y}\int\limits_{0}^{y}(u_{xx}(x,y))dydy.$\\
Following the algorithm (\ref{1}), the successive approximations are
\begin{eqnarray}
\begin{array}{l}\nonumber
u_{1}=N(u_{0})=-\frac{1}{2!}y^2g''(x)+\frac{1}{3!}y^3\cos(x)\\
u_{2}=N(u_{0}+u_{1})-N(u_{0})=\frac{1}{4!}y^4g^{iv}(x)+\frac{1}{5!}y^5\cos(x)\\
u_{3}=N(u_{0}+u_{1}+u_{2})-N(u_{0}+u_{1})=-\frac{1}{6!}y^6g^{vi}(x)+\frac{1}{7!}y^7\cos(x)\\
\vdots
\end{array}
\end{eqnarray}
Combining the above results, we obtain
\begin{equation}\label{6.3}
u(x,y)=\cos(x)\sinh(y)+g(x)-\frac{1}{2!}y^2g''(x)+\frac{1}{4!}y^4g^{iv}(x)-\frac{1}{6!}y^6g^{vi}(x)
\end{equation}
Using the inhomogeneous boundary condition, $u_y(x,\pi)=\cosh(\pi)\cos(x)$ , we obtain
$$\cos(x)\cosh(\pi)-\pi g''(x)+\frac{1}{3!!}\pi^3g^{iv}(x)-\frac{1}{6!}\pi^5 g^{vi}(x)+\cdots=\cosh(\pi)\cos(x)$$
Equating the coefficients of like terms on both sides gives
$$g''(x)=g^{iv}(x)=g^{vi}(x)=\cdots=0$$
which gives
$$g(x)=C_0$$
Consequently, the solution in series form is given by
\begin{equation}\nonumber
u(x,y)=\cos(x)(y+\frac{y^3}{3!}+\frac{y^5}{5!}+\cdots)+C_0
\end{equation}
and in closed form $$u(x,y)=\cos(x)\sinh(y)+C_0$$ which is also the exact solution.

\end{document}